\documentclass[12pt,preprint,a4paper]{aastex}
\usepackage{graphics,epsf}
\usepackage{amsmath}
\usepackage{amsfonts}
\usepackage{amssymb}
\usepackage{epsfig}
\usepackage{subfigure}
\usepackage{multirow}
\usepackage{longtable}
\usepackage{graphicx}

\def \kev{~\rm{keV}}

\def \cm{~\rm{cm}}
\def \s{~\rm{s}}
\def \km{~\rm{km}}

\def \K{~\rm{K}}
\def \g{~\rm{g}}

\def \erg{~\rm{erg}}

\def \yr{~\rm{yr}}
\def \Myr{~\rm{Myr}}

\def \kpc{~\rm{kpc}}

\def \keV{~\rm{keV}}

\altaffiltext{1}{Department of Physics, Technion -- Israel Institute of
Technology, Haifa 32000 Israel; agilkis@tx.technion.ac.il; soker@physics.technion.ac.il.}
\begin{document}

\title{HEATING THE INTRA-CLUSTER MEDIUM PERPENDICULAR TO THE JETS AXIS}
\author{Avishai Gilkis\altaffilmark{1} and Noam Soker\altaffilmark{1}}

\begin{abstract}
By simulating jet-inflated bubbles in cooling flows with the PLUTO hydrodynamic code we show that mixing of high entropy
shocked jet's material with the intra-cluster medium (ICM) is the major heating process perpendicular to the jets' axis.
Heating by the forward shock is not significant. The mixing is very efficient in heating the ICM in all directions,
to distances of $\sim 10 \kpc$ and more.
Although the jets are active for a time period of only $20 \Myr$, the mixing and heating near the equatorial plane, as well as along the symmetry axis,
continues to counter radiative cooling for times of $\ga 10^8 \yr$ after the jets have ceased to exist.
We discuss some possible implications of the results. ($i$) The vigorous mixing is expected to entangle magnetic field lines,
hence to suppress any global heat conduction in the ICM near the center.
($ii$) The vigorous mixing forms multi-phase ICM in the inner cluster regions, where the coolest parcels of gas will eventually cool first, flow inward, and feed the
active galactic nucleus to set the next jet-activity episode. This further supports the cold feedback mechanism.
($iii$) In cases where the medium outside the region of $r \sim 10 \kpc$ is not as dense as in groups and clusters of galaxies,
like during the process of galaxy formation, the forward shock and the high pressure of the shocked jets' material might expel gas from the system.
\end{abstract}

\section{INTRODUCTION}
\label{sec:intro}

The intra-cluster medium (ICM) in cooling flow (CF) clusters and groups of
galaxies is heated by a negative feedback mechanism (e.g., \citealt{Binney1995, Farage2012}),
mostly driven by active galactic nuclei (AGN) jets that inflate X-ray deficient cavities (bubbles; see, e.g.,
\citealt{Dong2010, OSullivan2011, Gaspari2012a, Gaspari2012b, Birzan2011, Gitti2012} for recent papers and
references therein).
Examples of bubbles include Abell 2052 \citep{Blanton2011}, NGC 6338 \citep{Pandge2012},
NGC 5044 \citep{David2009}, HCG 62 \citep{Gitti2010}, Hydra A \citep{Wise2007},
NGC 5846 \citep{Machacek2011} NGC 5813 \citep{Randall2011}, A 2597 \citep{McNamara2001},
Abell 4059 \citep{Heinz2002}, NGC 4636 \citep{Baldi2009},
NGC 5044 \citep{Gastaldello2009, David2011}, and RBS 797 \citep{Schindler2001, Cavagnolo2011, Doria2012}.

Wide bubbles very close to the origin of the jets (the AGN), e.g., as in
Abell 2052, that are termed `fat bubbles', can be inflated by jets that do
not penetrate through the ICM. Instead, they deposit their energy relatively
close to their origin and inflate the fat bubbles. Slow massive wide (SMW)
jets can inflate the fat bubbles that are observed in many CFs, in clusters,
groups of galaxies, and in elliptical galaxies \citep{Sternberg2007}. The
same basic physics that prevents wide jets from penetrating through the ICM
holds for precessing jets \citep{Sternberg2008a, Falceta-Goncalves2010}, or a
relative motion of the jets to the medium
\citep{Bruggen2007, Soker2009, Morsony2010, Mendygral2012}.
In the present study we will inflate bubbles by SMW jets, but our results
hold for bubbles inflated by precessing jets or a relative motion of the
ICM as well.
If the jets penetrate to a too large distance, then no bubbles are formed, while
in intermediate cases elongated and/or detached from the center bubbles
are formed (e.g., \citealt{Basson2003, Omma2004, Heinz2006, Vernaleo2006,
AlouaniBibi2007, Sternberg2007, ONeill2010, Mendygral2011, Mendygral2012}).

Vortices inside the bubbles and in their surroundings play major roles in
the formation of bubbles, their evolution, and their interaction with the
ICM (e.g. \citealt{Heinz2005}). \cite{Omma2004} find that a turbulent vortex
trails each cavity, and that this vortex contains a significant quantity of
entrained and uplifted material (also \citealt{Roediger2007}).
\cite{Sternberg2008b} find in their 2.5D numerical simulations that vortices
inside bubbles can stabilize them against the Rayleigh-Taylor (RT)
instability and can suppress the Kelvin-Helmholtz (KH) instability on the surface of each bubble.
Jet-excited shocks that interact with older bubbles can excite vortices that
dissipate energy to the ICM \citep{Friedman2012}. The vortices cause
semi-periodic changes in the bubble properties, such as its boundary.
This can lead a single bubble to excite several sound waves \citep{Sternberg2009}, and cause a single
jet episode to inflate a chain of bubbles \citep{Refaelovich2012}.
In the present study we further explore the role of vortices in the interaction of the jets and bubbles
with the ICM.
We concentrate on gas near the equatorial plane, i.e., ICM gas that does
not reside along the jets' expansion trajectory.

The heating of the gas perpendicular to the jets' axis need not be $100\%$
efficient, as observations show that heating does not completely offset
cooling (e.g., \citealp{Wis04, McN04, Cla04, Hic05, Bre06, Sal08, Wilman2009}),
and a \emph{moderate CF} exists \citep{Sok04}. \textit{Moderate} implies
here that the mass cooling rate to low temperatures is much lower than the
cooling rate expected without heating, but it is much larger than the
accretion rate onto the supermassive black hole (SMBH) at the center of the
cluster. The cooling mass is either forming stars (e.g., \citealp{Odea08, Raf08}),
forming cold clouds (e.g., \citealt{Edge2010}), accreted
by the SMBH to maintain the cold feedback mechanism \citep{Piz2010}, or is
expelled back to the ICM and heated, when it is shocked or mixed with the hot jets' material.

In the present paper we study the heating and expelling process of gas
residing near the equatorial plane. By running cylindrically symmetric
hydrodynamical simulations with the PLUTO code (section \ref{sec:numerical}) we study the flow
structure (section \ref{sec:flow}).
We then examine the degree of mixing and dredge-up of gas
from the equatorial plane vicinity (section \ref{sec:mixing}), and the heating processes
(section \ref{sec:heating}).
Our short summary is in section \ref{sec:summary}.

\section{NUMERICAL SET UP}
\label{sec:numerical}

The simulations were performed using the PLUTO code \citep{Mignone2007}. We use spherical
coordinates in 3D, but with an imposed azimuthal symmetry around the $\theta =0$ ($z$) axis.
Namely, a 2.5D grid where only the dependence on the $(r,\theta)$ coordinates is calculated.
We will present the results in a plane $\phi=$constant, the meridional plane, which we take to be
the $(\varpi,z)$ plane, where $\varpi$ is in the equatorial plane and $z$ is along the symmetry axis.
The grid radial and azimuthal domains are $r=0.5 - 215 \kpc$, and $\theta=0-90^\circ$, respectively. We focus on the relevant inner domain of $r \la 40 \kpc$; the large outer radius ensures that there are no boundary effects, at a low computational cost thanks to a radially stretching grid.

On the $r=0.5 \kpc$ inner boundary we enforce a jet outflow between angles $0-70^\circ$ for a limited time,
and reflective boundary conditions in the angular zone $70^\circ -90^\circ$.
In the simulations presented here the two jets power is $P_{2j} = 2 \times 10^{44} \erg \s^{-1}$, their initial
velocity is $v_j = 9600 \km \s^{-1}$, and mass loss rate into the two jets is $\dot M_{2j} = 7 M_\odot \yr^{-1}$.
Such massive wide sub-relativistic outflows are supported by recent observations (e.g., \citealt{Moe2009, Dunn2010, Tombesi2012}).
At $t=20\Myr$ we turn the jet off, and apply reflective boundary conditions on the entire inner sphere.
We set reflective boundary conditions on the boundaries $\theta =0$ ($z$ axis) and $\theta =90^\circ$ (the equatorial plane).
The latter boundary condition mimics a symmetric opposite jet.
The initial density profile has a spherical symmetry, with a profile of the form (e.g., \citealp{Vernaleo2006})
\begin{equation}
\rho (r) =\rho_{c} \left[ 1+\left( \frac{r}{r_{0}} \right)^{2} \right] ^{-\frac{3}{4}}.
\label{eq:rho1}
\end{equation}
We take here $r_0 = 100 \kpc$, and $\rho_{c} = 10^{-25} \g \cm^{-3}$.

We start the simulation with an isothermal sphere at temperature $T=4 \times 10^7 \K$,
so that the pressure profile is proportional to the
density profile $p=c^2 \rho / \gamma$, where $c=(\gamma p / \rho)^{1/2} = (\gamma kT / \mu m_p)^{1/2}$
is the sound speed, with symbols having their usual meaning, and $\gamma =\frac{5}{3}$.
Radiative cooling is included by using table 6 from \cite{Sutherland1993}.
We use a time invariable spherical gravity field calculated
from the hydrostatic equilibrium at $t=0$
\begin{equation}
g(r)= \frac{c^2}{\gamma}\frac{1}{\rho}\frac{d\rho}{dr}=
-\frac{3}{2}\frac{c^2}{\gamma}\frac{r}{r_{0}^2}\left[1+\left(\frac{r}{r_{0}}\right)^2\right]^{-1}.
\label{eq:gravity1}
\end{equation}

To quantitatively analyze the thermal evolution of the ICM we marked
several regions of the initial ambient gas in the $(\varpi,z)$ plane. Each such region is actually a torus
due to our 2.5D grid.
These are called the {\it traced regions}.
The tracing is done by defining artificial flow quantities called 'tracers' in the PLUTO code \citep{Mignone2007},
that are frozen-in to the flow.
At $t=0$ the tracer of the region centered at $(\varpi,z)=(a,b)$ and having a radius of $0.25 \kpc$ is set to $\xi=1$.
Mixing of the traced gas with the ICM or the jet's material changes the tracer to values below $1$.
Since this quantity is advected with mass, the summation over all zones $i$ of mass multiplied by tracer value
$M_i \times \xi_i$ is constant with time.
This was verified in the post-simulation analysis.
Using these tracers we define the average property Q of the traced gas that was centered on $(\varpi,z)=(a,b)$ at $t=0$
\begin{equation}
Q_{ab} \equiv \frac{\Sigma_i \xi_i M_i Q_i}{\Sigma_i \xi_i M_i},
\label{eq:tracers1}
\end{equation}
where $Q$ can be the temperature, the pressure, or the position of the traced gas,
i.e., its center of mass.

\section{THE GENERAL FLOW STRUCTURE}
\label{sec:flow}

\begin{figure}
\hskip -1.0 cm
\begin{tabular}{cc}
\hskip 0.55 cm
{\includegraphics*[scale=0.7, trim=4.2cm 0 3.9cm 0]{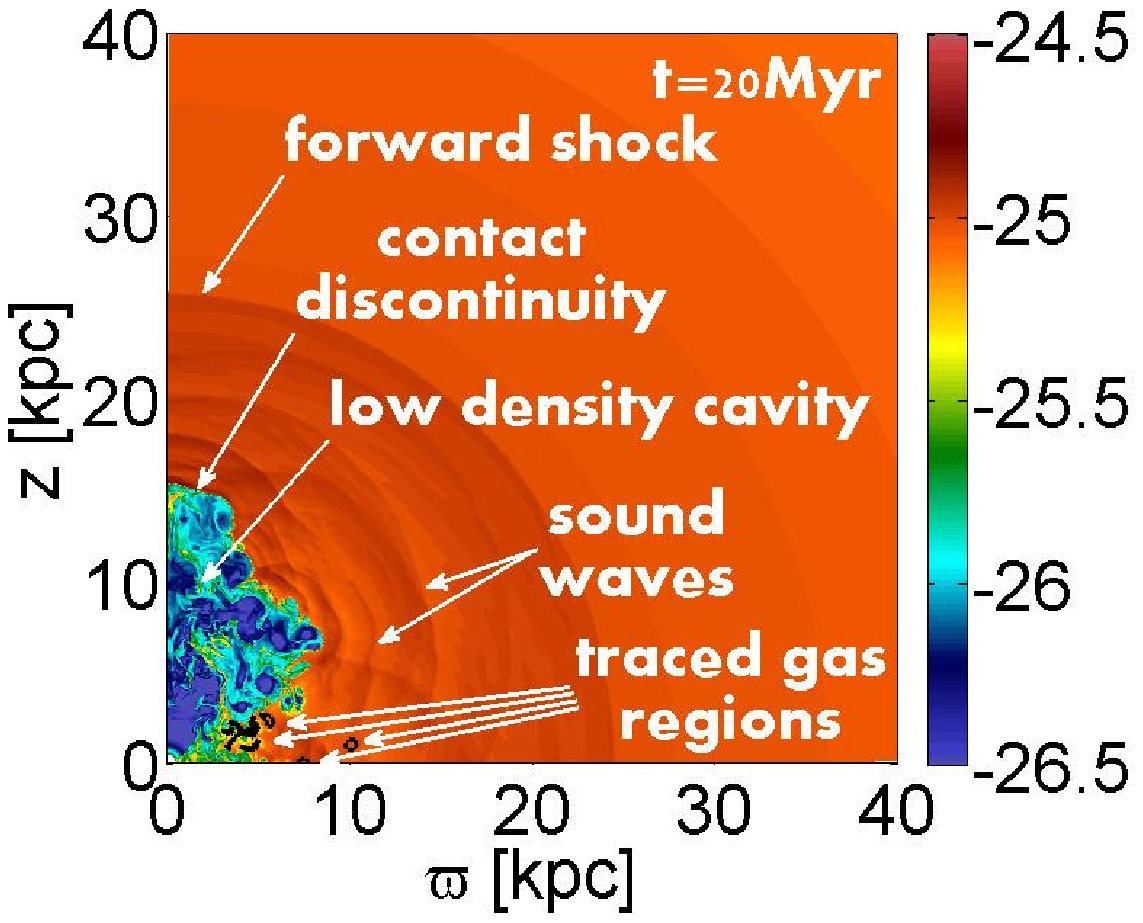}} &
\hskip +0. cm
{\includegraphics*[scale=0.7, trim=4.2cm 0 3.9cm 0]{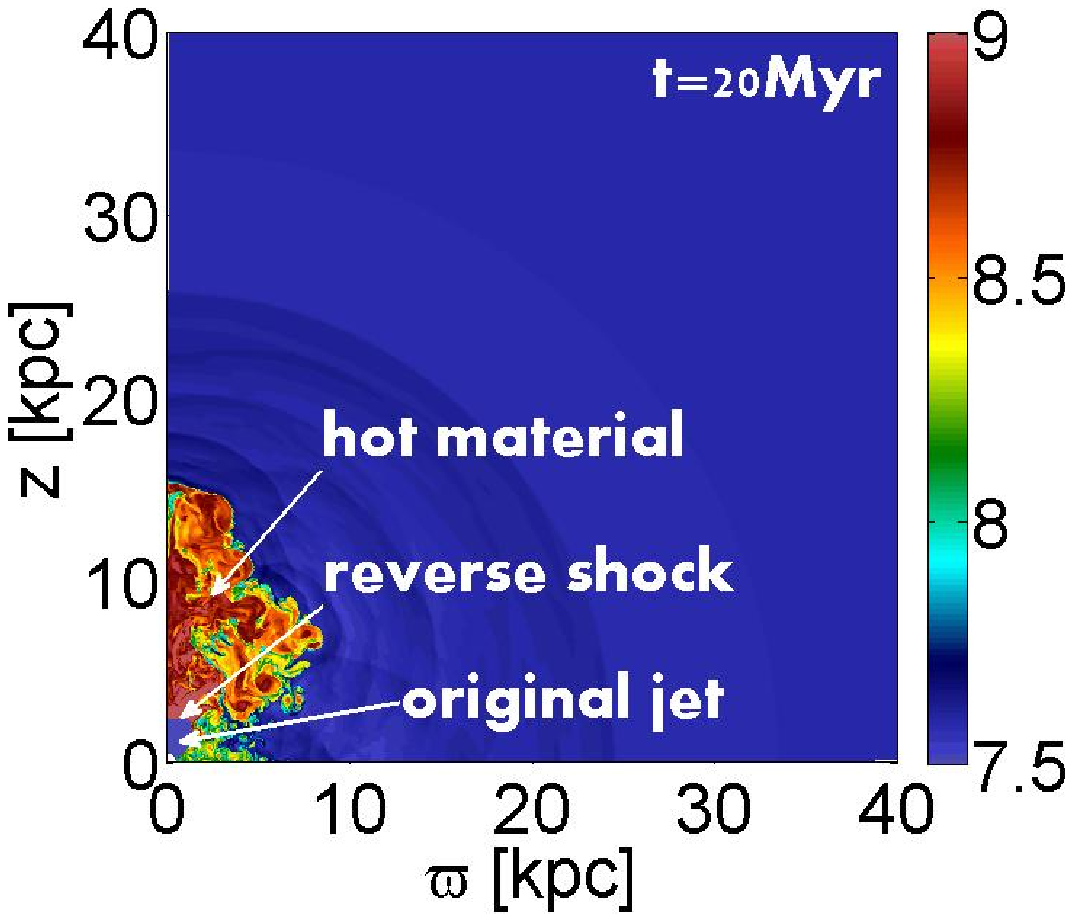}} \\
\end{tabular}
      \caption{The flow structure at $t=20\Myr$ in the $(\varpi,z)$, where $z$ is along the symmetry (jet) axis
      and $\varpi$ is in the equatorial plane.
      The simulation is 3D with axisymmetry imposed, namely, a 2.5D simulation.
      Some prominent features of the flow are marked.
      Left: density map with the traced regions marked
      (TR31, TR42, TR60 and TR91).
      Due to our 2.5D grid, each traced region is a thin torus.
      Density scale is in units of $\log \rho({\rm g} \cm^{-3}).$
       Right: the temperature map in units of $\log T(\K)$. }
      \label{fig:presentation}
\end{figure}

\begin{figure}
  \centering
    {\includegraphics[width=0.7\textwidth]{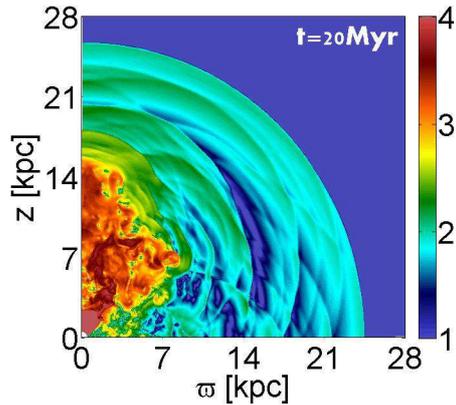}}
      \caption{The velocity map in the $(\varpi,z)$ plane at $t=20 \Myr$.
       The forward shock, reverse shock, and particularly the sound waves, are clearly visible.
      Velocity scale is in units of $\log v({\rm km} \s^{-1})$.
      }
      \label{fig:velocity2}
\end{figure}

In figure \ref{fig:presentation} we show the density (left) and temperature (right) maps at the end of the jet's injection phase ($t=20\Myr$).
In this simulation the combined two jets (here we simulate only one jet) power is $P_{2j} = 2 \times 10^{44} \erg \s^{-1}$, their initial
velocity is $v_j = 9600 \km \s^{-1}$, and mass loss rate into the two jets is $\dot M_{2j} = 7 M_\odot \yr^{-1}$.
Clearly identified in the figure are the forward shock running into the ICM, multiple sound waves
(clearly seen in the velocity map presented in figure \ref{fig:velocity2}),
and the reverse shock, where the jet is shocked.
The shocked jet material forms a hot low density bubble, of size $\sim 10 \kpc$ at $t=20\Myr$.
The bubble is the region of low density (dark blue and light blue) in figure
\ref{fig:presentation} (left), corresponding to the hot regions (orange, red and yellow)
in figure \ref{fig:presentation} (right).
The bubble obtained here is similar to the results of \cite{Sternberg2007}.
Such bubbles are observed as X-ray deficient bubbles (cavities).
Also marked is the contact discontinuity separating the shocked jet's material from the shocked ICM gas.
At later times mixing prevails and the contact discontinuity cannot be identified (see next section).
The low post-shock velocity of $85 \km \s^{-1}$ in the equatorial plane is because the forward
shock in the equatorial plane at $t=20 \Myr$ is very weak, having a Mach number of $M_{\rm eq} (20) = 1.1$.

\begin{figure}
  \centering
    {\includegraphics[width=0.7\textwidth]{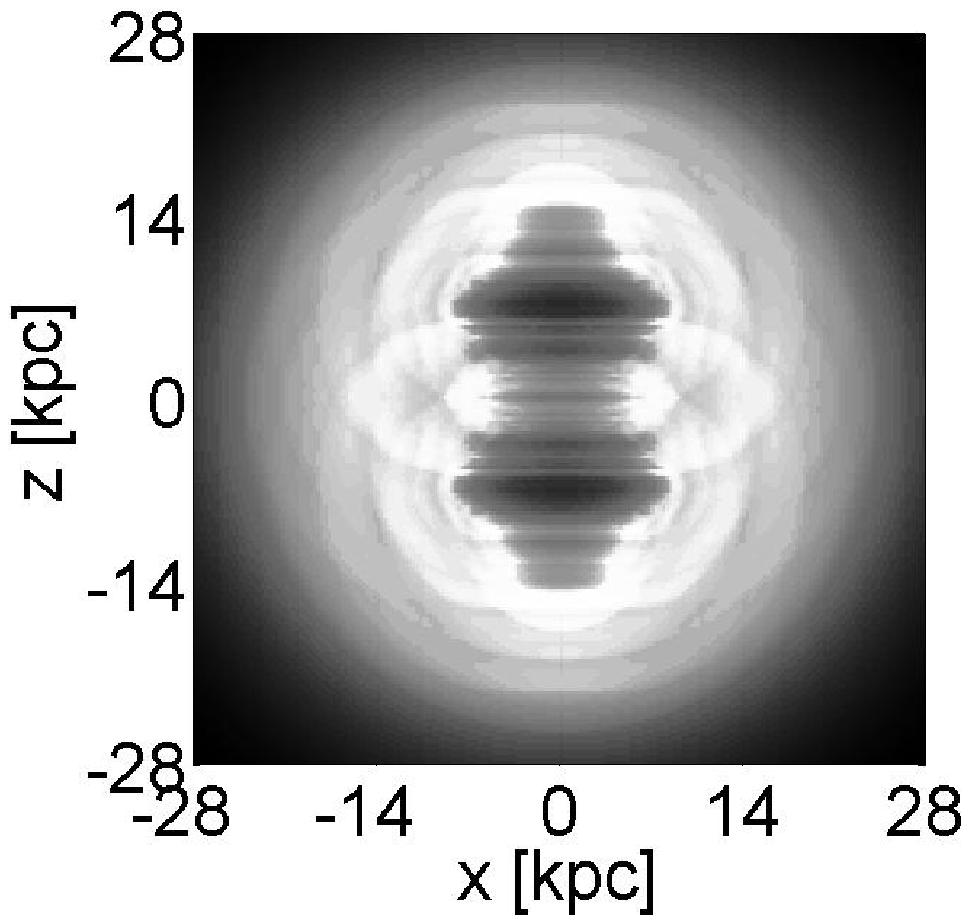}}
      \caption{The projected X-ray map in the $(x,z)$ plane at $t=20 \Myr$, obtained by
      integrating density squared along the line of sight.
      The $x$ axis on the plane of the sky coincides with the $\varpi$ axis of figure \ref{fig:presentation}.
      Only one quarter of the plane was simulated, and it was mirrored twice to obtain the full image
      presented here.
      The bubbles are clearly seen as regions of low-intensity  emission.
      Also clearly seen are sound waves. 
      }
      \label{fig:xrayim}
\end{figure}

To demonstrate the similarity to observed X-ray deficient bubbles, we produce a synthetic X-ray image by
integrating the density squared along the line of sight perpendicular to the $(x,z)$ plane,
where the $x$ axis on the plane of the sky coincides with the $\varpi$ axis of figure \ref{fig:presentation}.
A simulated X-ray image of the cluster inner region at $t=20 \Myr$ is presented in figure
\ref{fig:xrayim}.

\begin{figure}
  \centering
    {\includegraphics[width=0.7\textwidth]{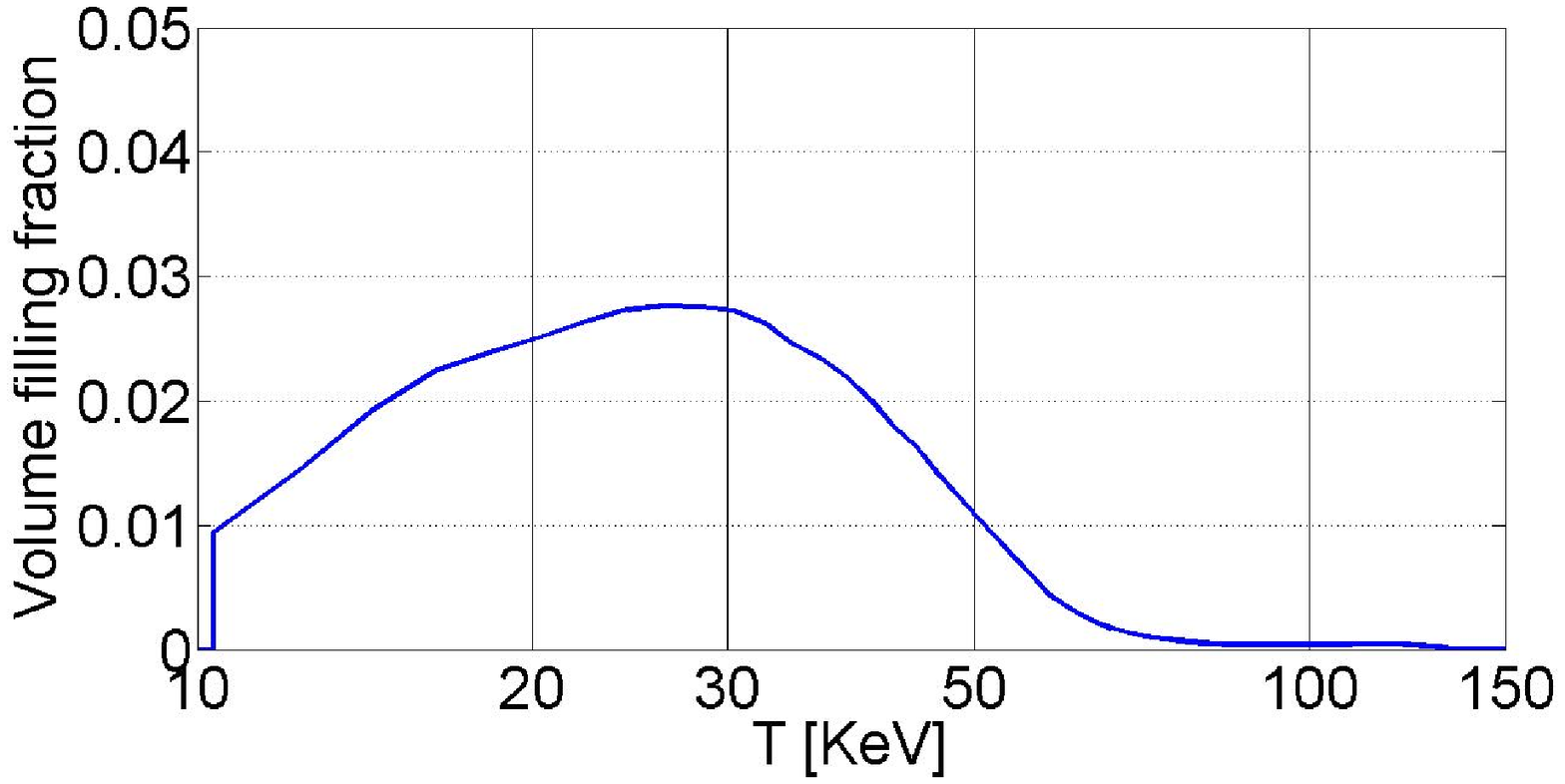}}
      \caption{The volume filling fraction (per \keV) of the bubble as
      function of temperature. The bubble here was defined as material
      with temperature above $3$ times the initial ambient temperature.
      }
      \label{fig:vfillingf}
\end{figure}

The thermal content of the bubble at $t=20 \Myr$ is presented as filling factor (per $\keV$) as function of temperature in figure \ref{fig:vfillingf}.
The bubble here was defined as material having a temperature $T_b>3T=10.3 \kev$, where $T=4 \times 10^7 \K$ is the initial ICM temperature.
At $t=20 \Myr$ the bubble volume is equal to that of a sphere of radius $\sim 7 \kpc$.
The volume changes a little for small changes of the temperature threshold.
The post shock temperature of the jet (for $\gamma=5/3$ and $v_j = 9600 \km \s^{-1}$ used here) is $T_{pj}=110 \kev$.
The temperature of the bubble is much lower, as energy is transferred to the ICM, by performing work on the ICM to inflate the bubble
and by heating ICM gas that is mixed into the bubble (see section \ref{sec:mixing}).
The resulting volume filling fraction is well below the upper limit set by \cite{Sanders2007}
for the Perseus cluster, although the work presented here is general
and we have not set out to reproduce a specific observed system.

We pick several thin tori at $t=0$ and follow the evolution of the material inside them.
These are called the {\it traced regions}. The four traced regions marked on
figure \ref{fig:presentation} were located initially at $(\varpi,z)_0=(3,1)$,
$(4,2)$,
$(6,0)$, and $(9,1)$; these are called TR31,
TR42,
TR60 and TR91, respectively.
By $t=20\Myr$ they moved to $(\varpi,z) \simeq (5,1)$,
$(5.5,2.5)$,
$(7.5,0)$, and $(10,1)$, respectively.
We note that TR31
and TR42 were
pushed outward and lost
their
circular cross section due to interaction with the shocked gas,
both from the jet and the ICM.
TR60 and TR91 can be seen to have moved a more-or-less uniform distance of $1\kpc$ away from the jet axis.
Their motion was set by the forward shock.

\begin{figure}
\hskip -1.0 cm
\begin{tabular}{cc}
\hskip 0.55 cm
{\includegraphics*[scale=0.7, trim=4.2cm 0 3.9cm 0]{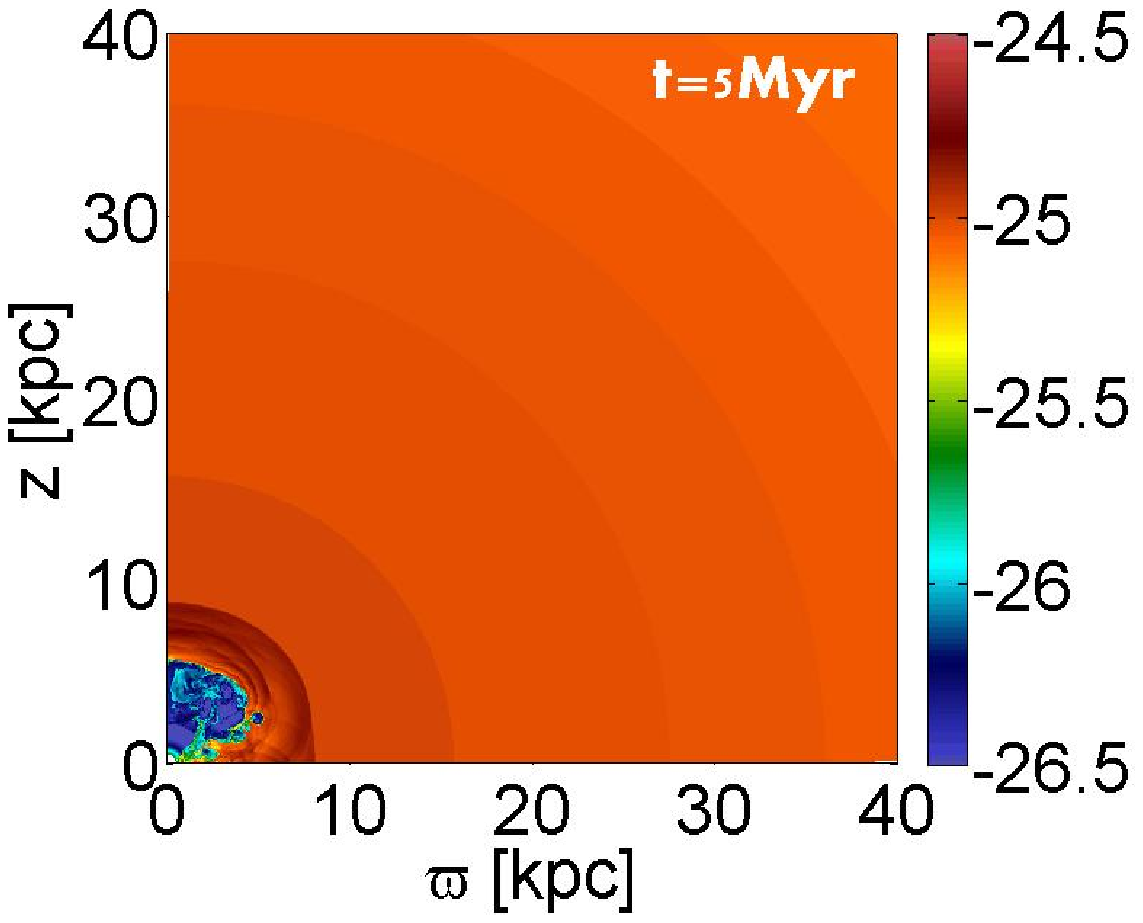}} &
\hskip -0.39 cm
{\includegraphics*[scale=0.7, trim=4.2cm 0 3.90cm 0]{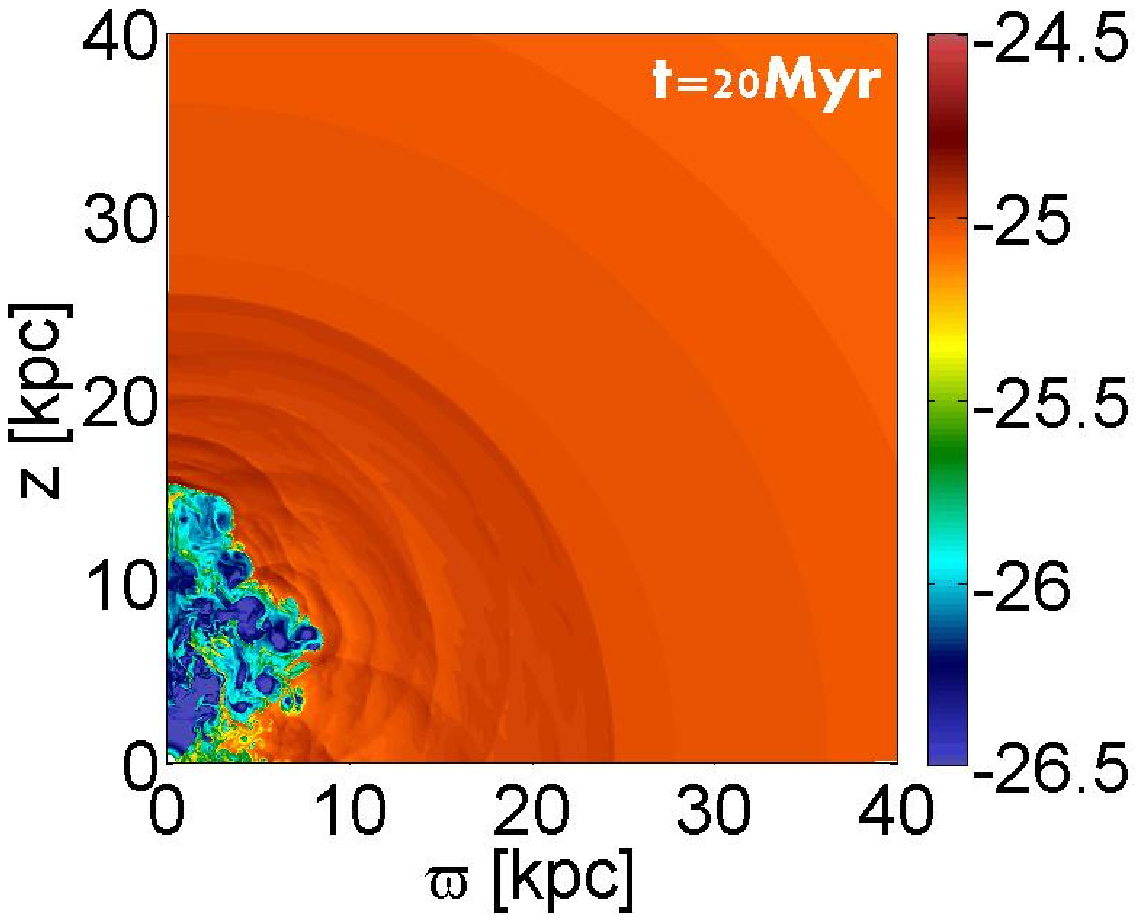}} \\
\hskip 0.55 cm
{\includegraphics*[scale=0.7, trim=4.2cm 0 3.9cm 0]{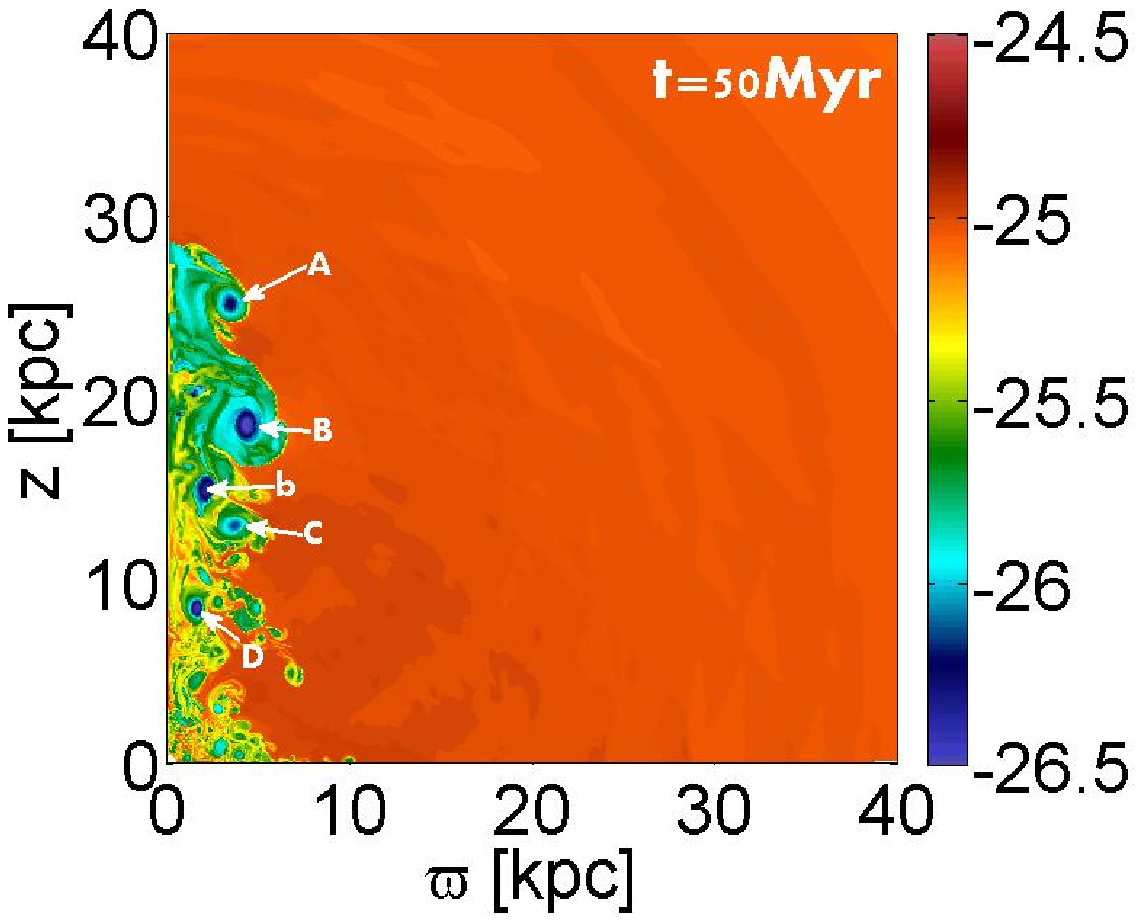}} &
\hskip -0.39 cm
{\includegraphics*[scale=0.7, trim=4.2cm 0 3.9cm 0]{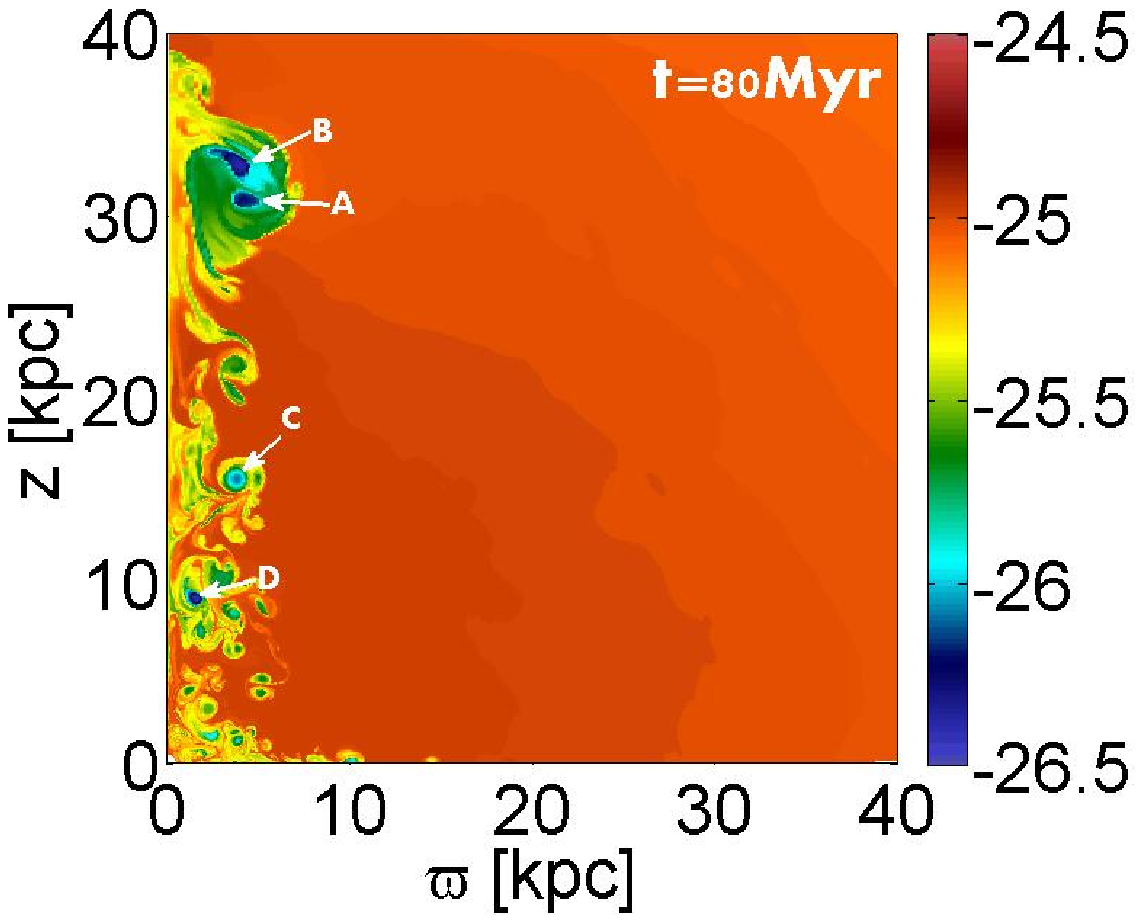}} \\
\end{tabular}
      \caption{The density maps in the $(\varpi,z)$ plane at four times: $t=5$, $20$, $50$, and $80 \Myr$.
      The jet is active during the $t=0-20 \Myr$ time period. Density scale is in units of $\log \rho(\g \cm^{-3})$.
      We follow the position of five vortices and mark them on the panels at $t=50$ and $80 \Myr$.
      Note that vortices B and b merge in the time lapse between $t=50$ and $80 \Myr$. }
      \label{fig:evolution}
\end{figure}

\begin{figure}
\hskip -1.0 cm
\begin{tabular}{cc}
\hskip 0.55 cm
{\includegraphics*[scale=0.7, trim=4.2cm 0 4.2cm 0]{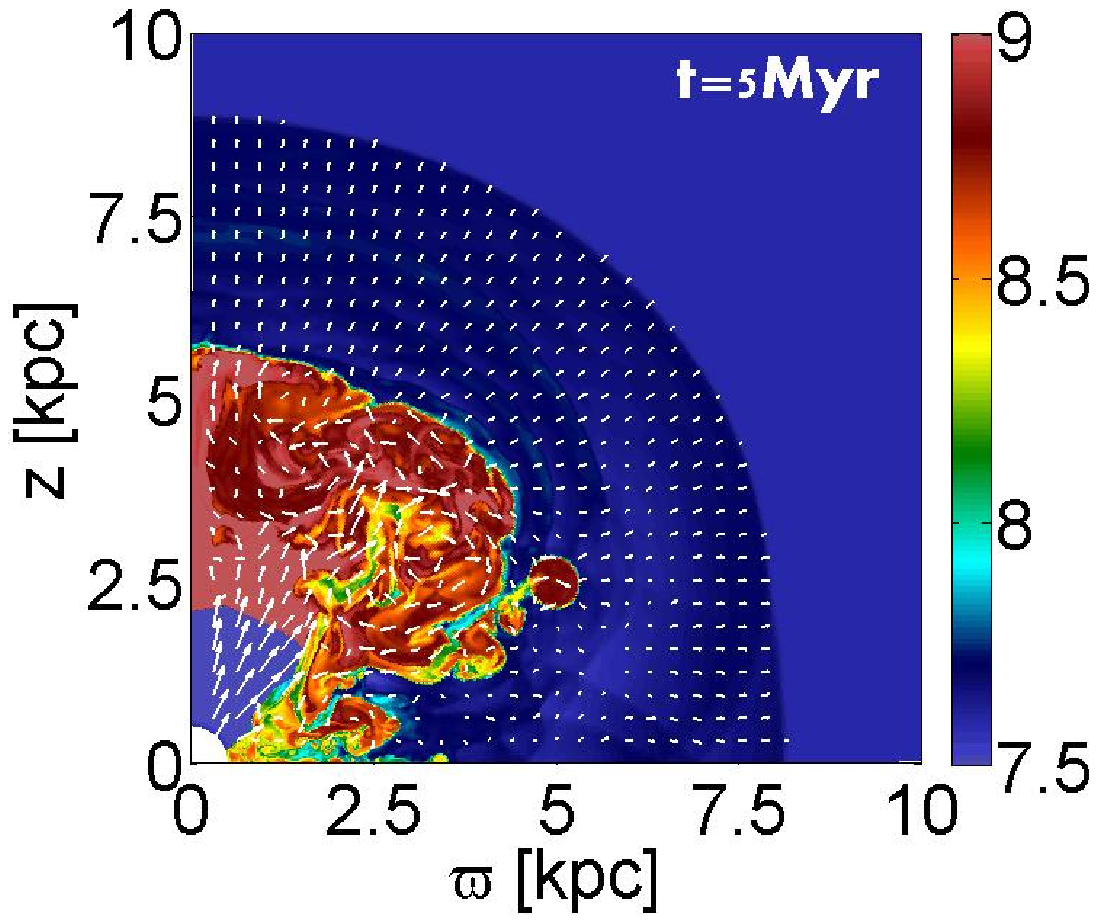}} &
\hskip -0.25 cm
{\includegraphics*[scale=0.7, trim=4.2cm 0 4.2cm 0]{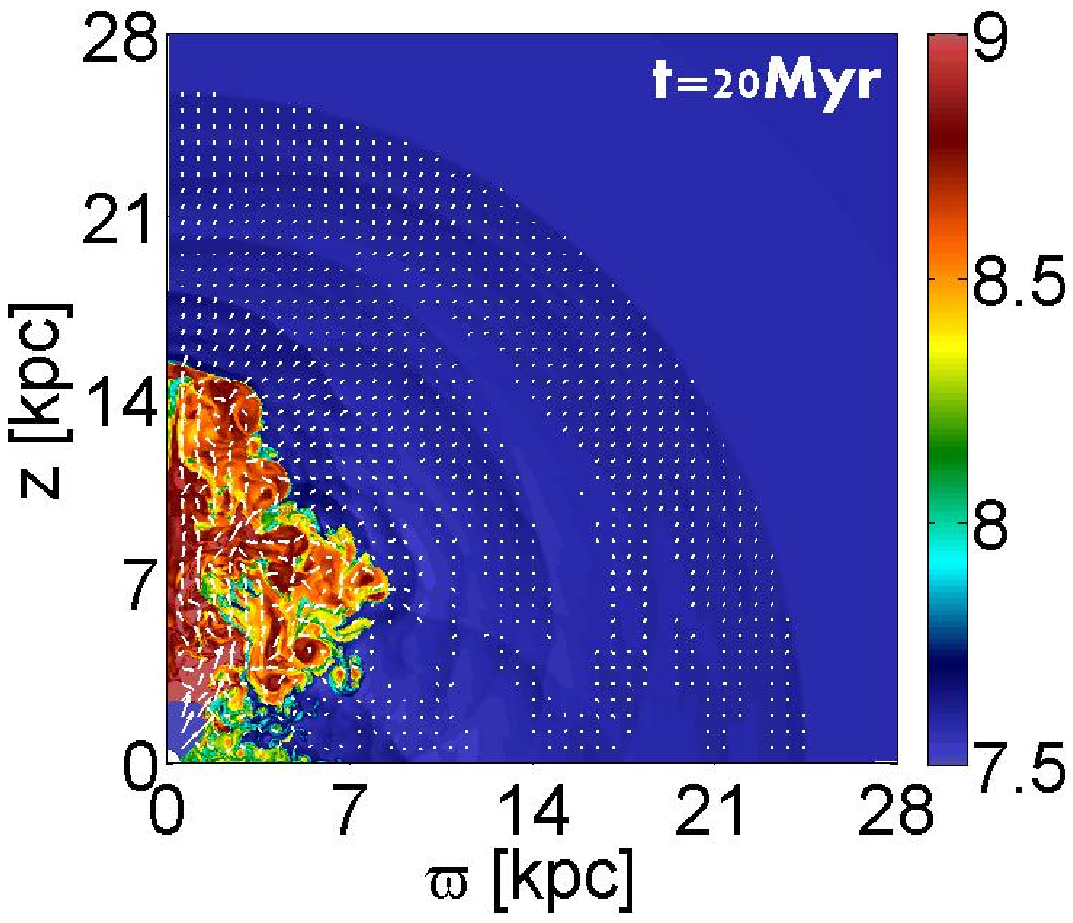}} \\
\hskip 0.55 cm
{\includegraphics*[scale=0.7, trim=4.2cm 0 4.2cm 0]{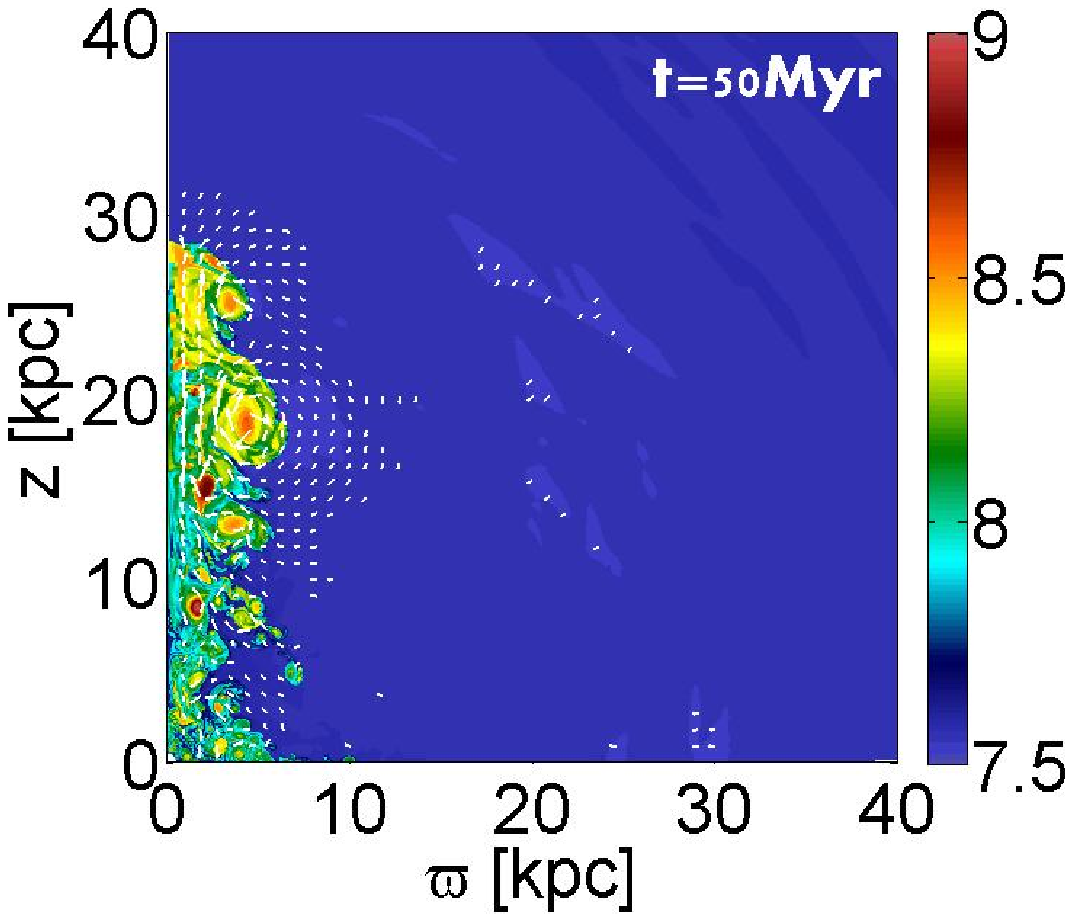}} &
\hskip -0.25 cm
{\includegraphics*[scale=0.7, trim=4.2cm 0 4.2cm 0]{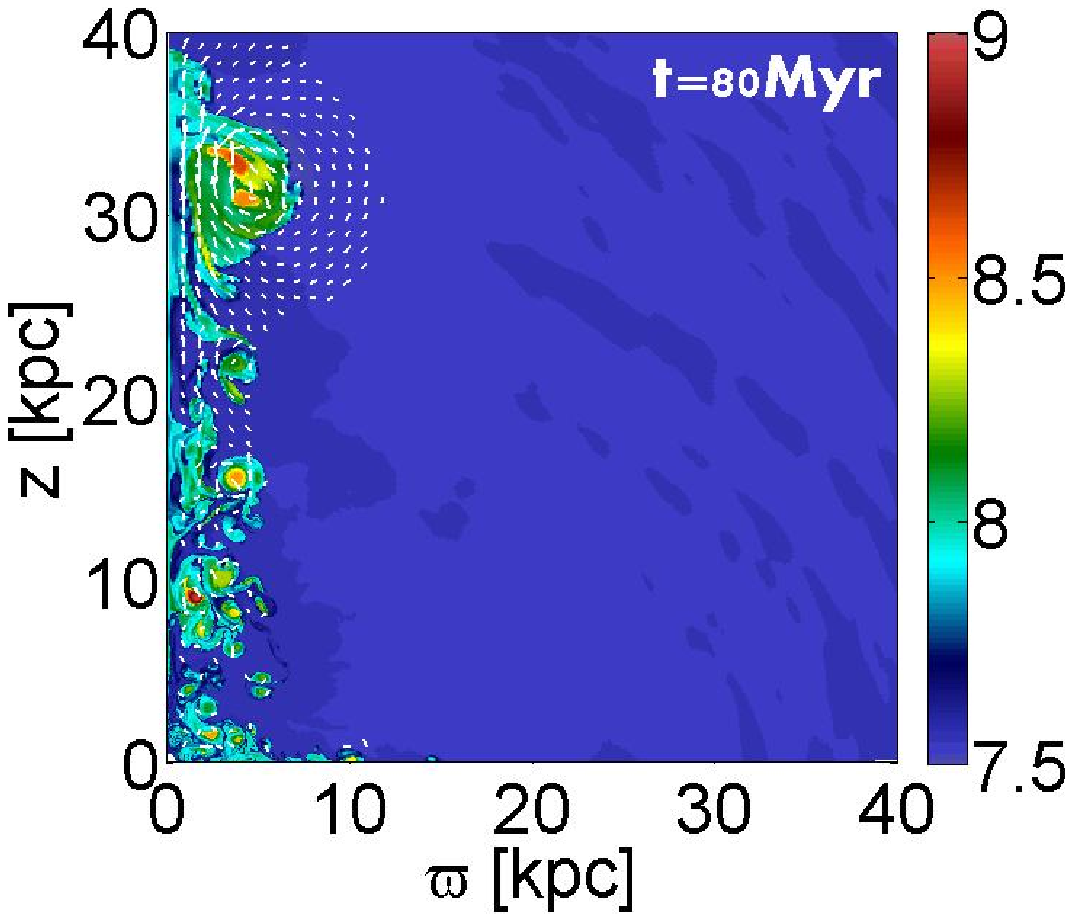}} \\
\end{tabular}
      \caption{Like figure \ref{fig:evolution} but for the velocity and temperature maps.
      A large and clear vortex, centered at $(\varpi,z)_{\rm v} \simeq(2.6,2.6)\kpc$,  is seen in the shocked jet's gas at $t=5 \Myr$.
       The temperature scale is in units of $\log T(\K)$, while the velocity vectors are divided into four groups by their length,
       from longest to shortest in ${\rm km} \s^{-1}$: $5000 < v$, $1000< v \le 5000$, $200<v \le 1000$, and $50<v \le 200$.}
   \label{fig:velocity}
\end{figure}

Figure \ref{fig:evolution} shows the evolution of the flow structure.
Already at $t=5\Myr$ a clear bubble is seen.
At $t=50\Myr$ the bubble has moved away from the center by inertia and buoyancy.
Also seen is a trailing region of vortices that drags ICM outward.
By $t=80\Myr$ we can still identify the bubble, although it starts to dissipate.
The velocity maps at these times are presented in figure \ref{fig:velocity}.
Most of the bubble's volume is filled with a large vortex.
In our 2.5D simulations the vortex has a shape of a torus, but in a realistic 3D flow we expect the presence of a more complex
structure of vortices.
The evolution of vortices, inside and outside the bubbles, play a major role in the heating and mixing as we
show in the next section. Here we follow the position of five vortices from $t=50\Myr$ to $t=80\Myr$ as marked on
figure \ref{fig:evolution}. Vortices A, B and b merge, vortex C slowly moves outward, while vortex D stays at about the same place.
At $t=80 \Myr$ the large vortex A, center at $(\varpi,z)_v=(4,30) \kpc$, includes both the shocked jet material (close to the symmetry axis) and the ICM.
This holds also to the C and D vortices, centered on $(\varpi,z)_v=(4,15.5)\kpc$ and $(\varpi,z)_v=(1.5,9)\kpc$, respectively.
\cite{Refaelovich2012} further discuss the evolution of vortices and their role in determining the morphology and evolution of bubbles
in groups and clusters of galaxies (for early studies of vortices in general contexts see \citealt{Norman1996} and references therein).

\section{MIXING AND DREDGE UP}
\label{sec:mixing}

In the previous section the inflation of `fat bubbles' and the significant role of vortices in the evolution of bubbles were demonstrated
(see also \citealt{Sternberg2007, Sternberg2008a, Sternberg2008b, Refaelovich2012}).
We now turn to examine the mixing process.

\begin{figure}
\hskip -1.0 cm
\begin{tabular}{cc}
\hskip 0.55 cm
{\includegraphics*[scale=0.7, trim=4.2cm 0 4.2cm 0]{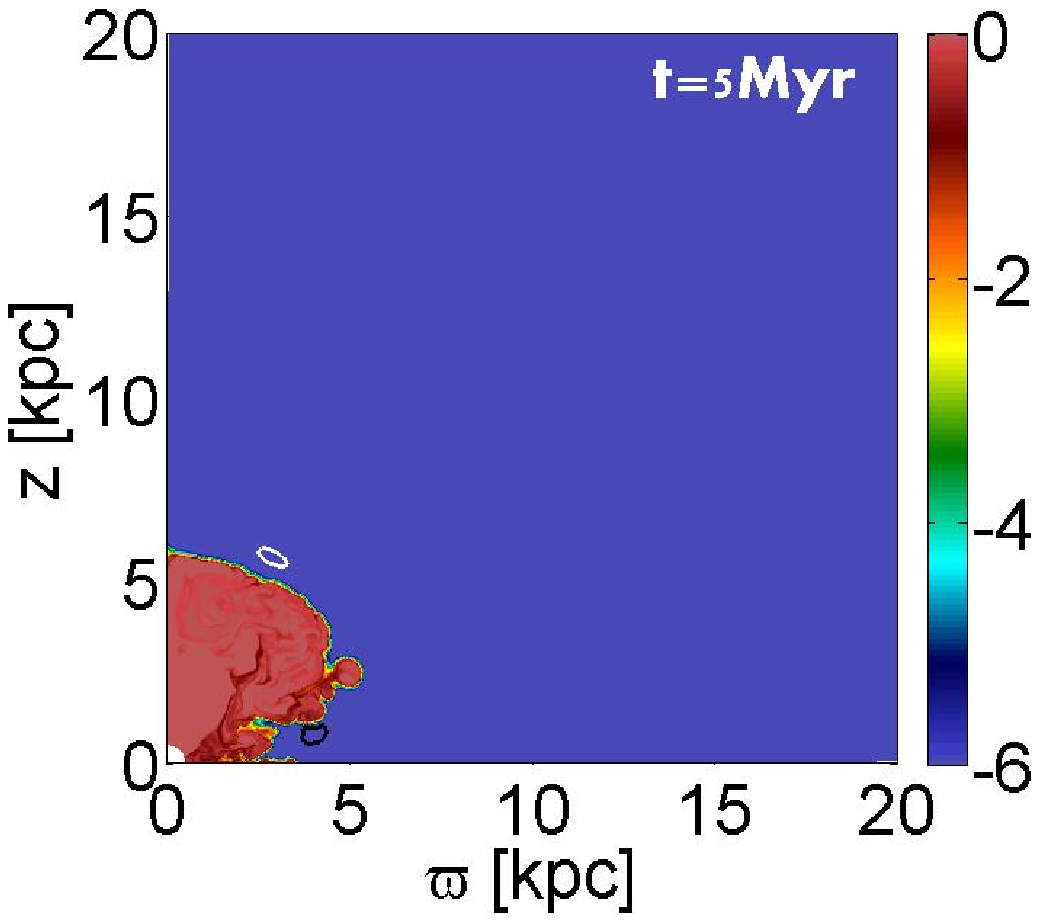}} &
\hskip -0.25 cm
{\includegraphics*[scale=0.7, trim=4.2cm 0 4.2cm 0]{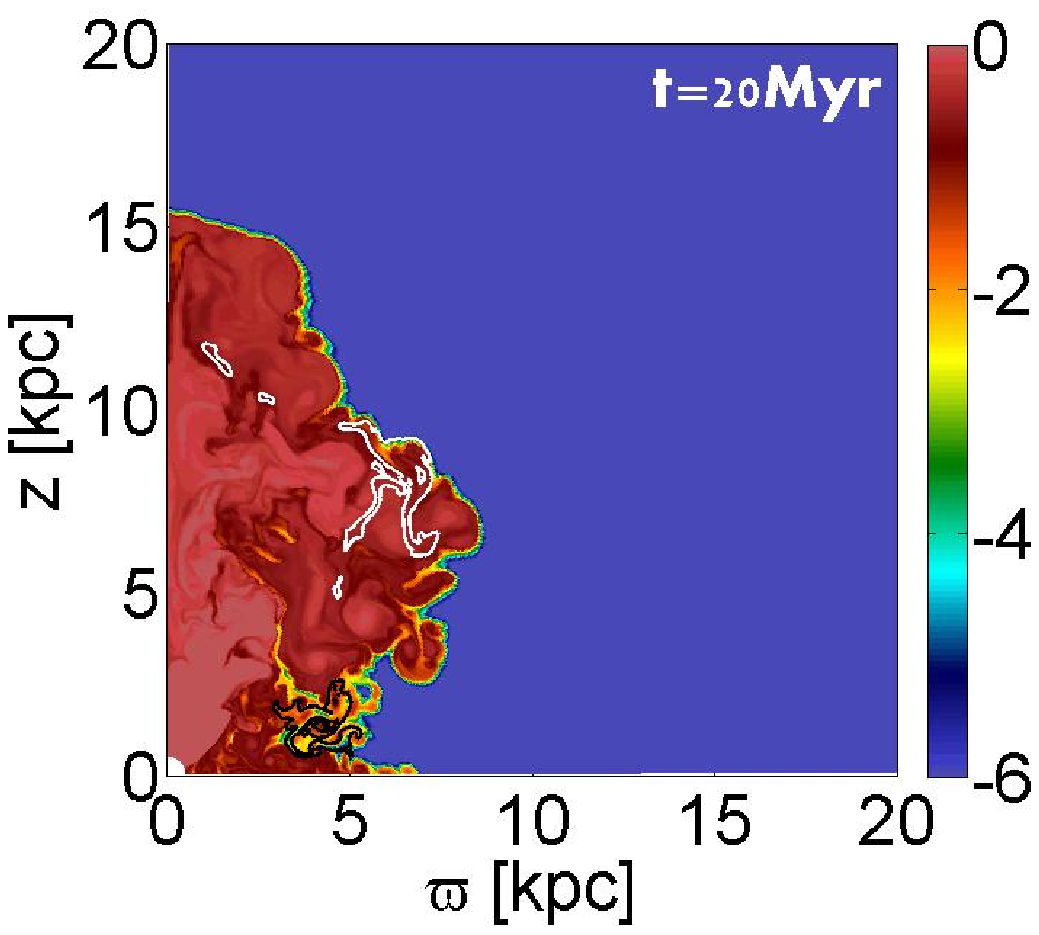}} \\
\hskip 0.55 cm
{\includegraphics*[scale=0.7, trim=4.2cm 0 4.2cm 0]{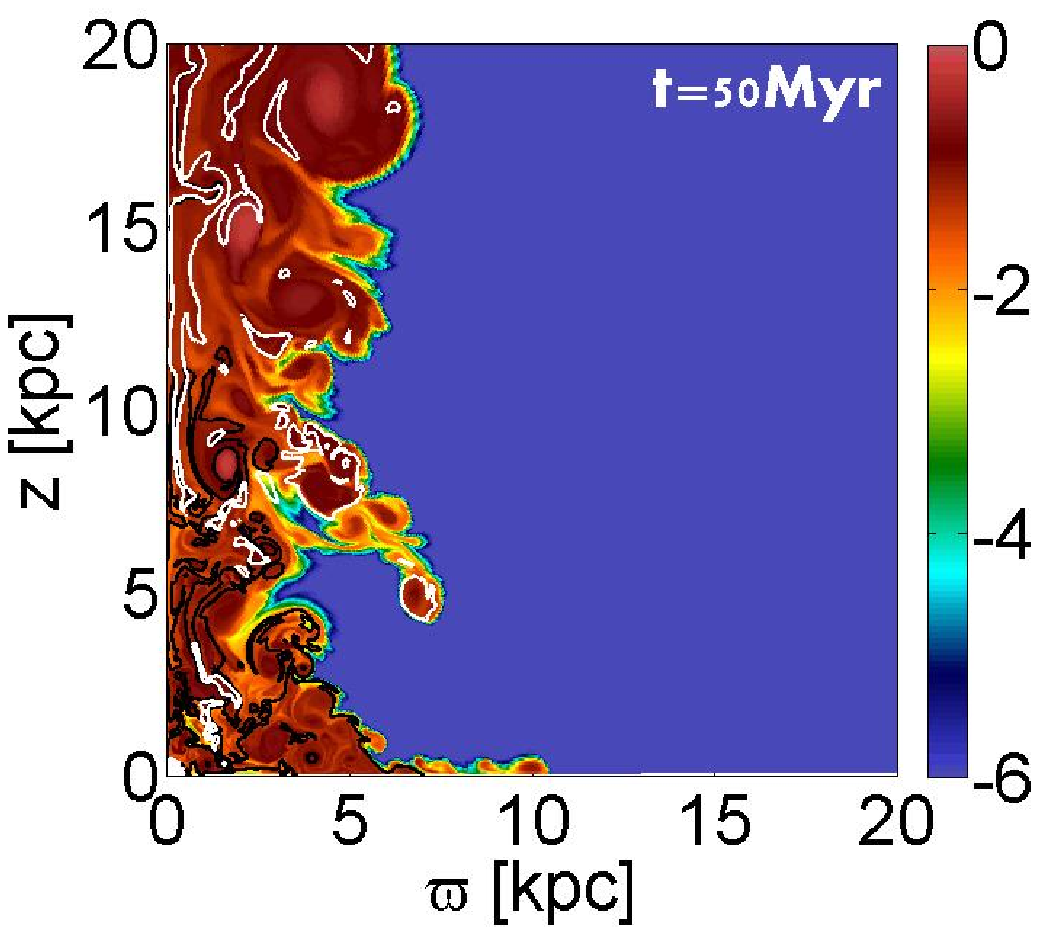}} &
\hskip -0.25 cm
{\includegraphics*[scale=0.7, trim=4.2cm 0 4.2cm 0]{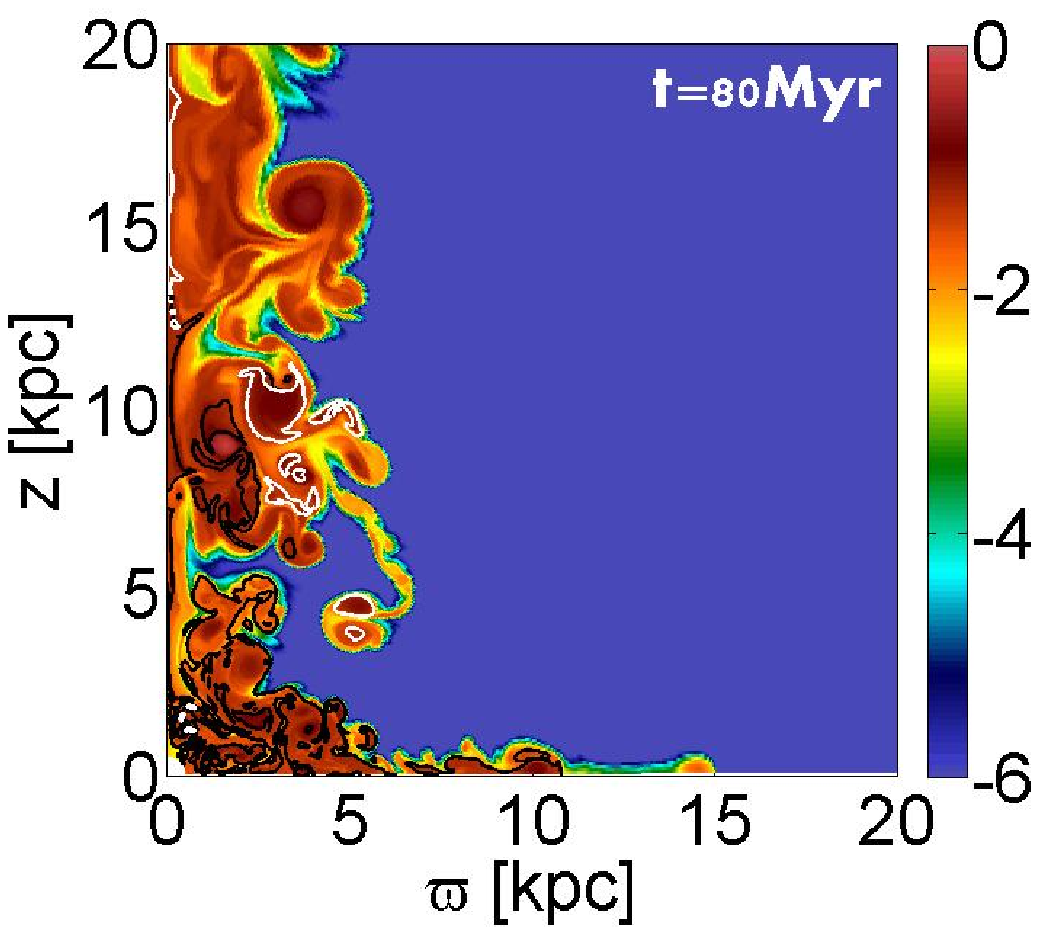}} \\
\end{tabular}
      \caption{The concentration of jet material $\xi_{\rm jet}$ (fraction of jet's material at each point) is shown by the color
      coded map in logarithmic scale at four times.
      It is clearly seen that the jet's material mixes with the ICM, and as shown in the next section, heats it.
      We also follow the material of two tracers, TR31 and TR24.
      Here TR31 is the gas that started inside the circular region centered on $(\varpi, z)=(3,1)\kpc$ and having a radius of $0.25\kpc$,
      with a similar definition for  TR24.
      The black (white) contours show where the concentration of TR31 (TR24) is one percent.
      Most of the TR24 gas is carried with the jet's gas along the symmetry axis.
      The TR31 gas suffers vigorous mixing near the center, and is heated up by this mixing.}
      \label{fig:mixing1}
\end{figure}

We first follow the mixing of the jet's material with the ICM and with the traced region TR24,
i.e., the traced region whose cross section
in the $(\varpi, z)$ plane at $t=0$ is a circle of radius $0.25 \kpc$ centered on $(\varpi, z)=(2,4)$, as well as with the TR31 tracer.
Figure \ref{fig:mixing1} shows the jet's material in colors, the TR24 gas in white contours,
and the TR31 gas in black contours; contours show where the concentration of each tracer is
one percent.
The color code gives the fraction of the original jet's material at each point.
At early times, $t \la 10 \Myr$, the material near the axis is almost purely of jet origin (light red).
Further away from the axis, where the shocked jet's gas resides, the so called `cocoon', gas from the ICM
is mixed with the shocked jet material, as can be seen by the dark red regions.
At $t=20 \Myr$ the shocked jet's gas that fills most of the bubble is heavily mixed with the ICM.
At $t=20 \Myr$ the jets cease, and a region trailing the main (front) bubbles is formed.
At $t=50 \Myr$ the main bubble is on the upper left, while at $t=80 \Myr$ the main bubble is outside the frame of the lower-right
panel of figure \ref{fig:mixing1}  (see fig. \ref{fig:evolution}).
At $t= 50 \Myr$ a substantial mixing is seen in the trailing volume, as indicated by the yellow and orange colors.
This mixing is more significant at $t=80 \Myr$. Pockets of ICM gas can be seen within these mixed regions at these two times.
White contours that mark the position of the TR24 traced region clearly demonstrate the mixing and dredge-up
of the ICM.
Mixing seems to a be the major process of heating the ICM along the jet's propagation direction.

We now turn to the mixing of gas perpendicular to the jets' axis.
At early times (here $t=5\Myr$) the traced region TR31 has been pushed by the forward shock and the
pressure of the shocked jet's material, while still staying intact.
At later times the TR31 initial morphology (cross section in the $\varpi, z$ plane) is violently disrupted
by the turbulence (vortices) of the shocked jet and ICM gas.
By $t=80 \Myr$ the original TR31 gas is heavily mixed with the hot shocked gas, and its temperature increases as we show in the
next section.

\begin{figure}
  \centering
    {\includegraphics[width=0.7\textwidth]{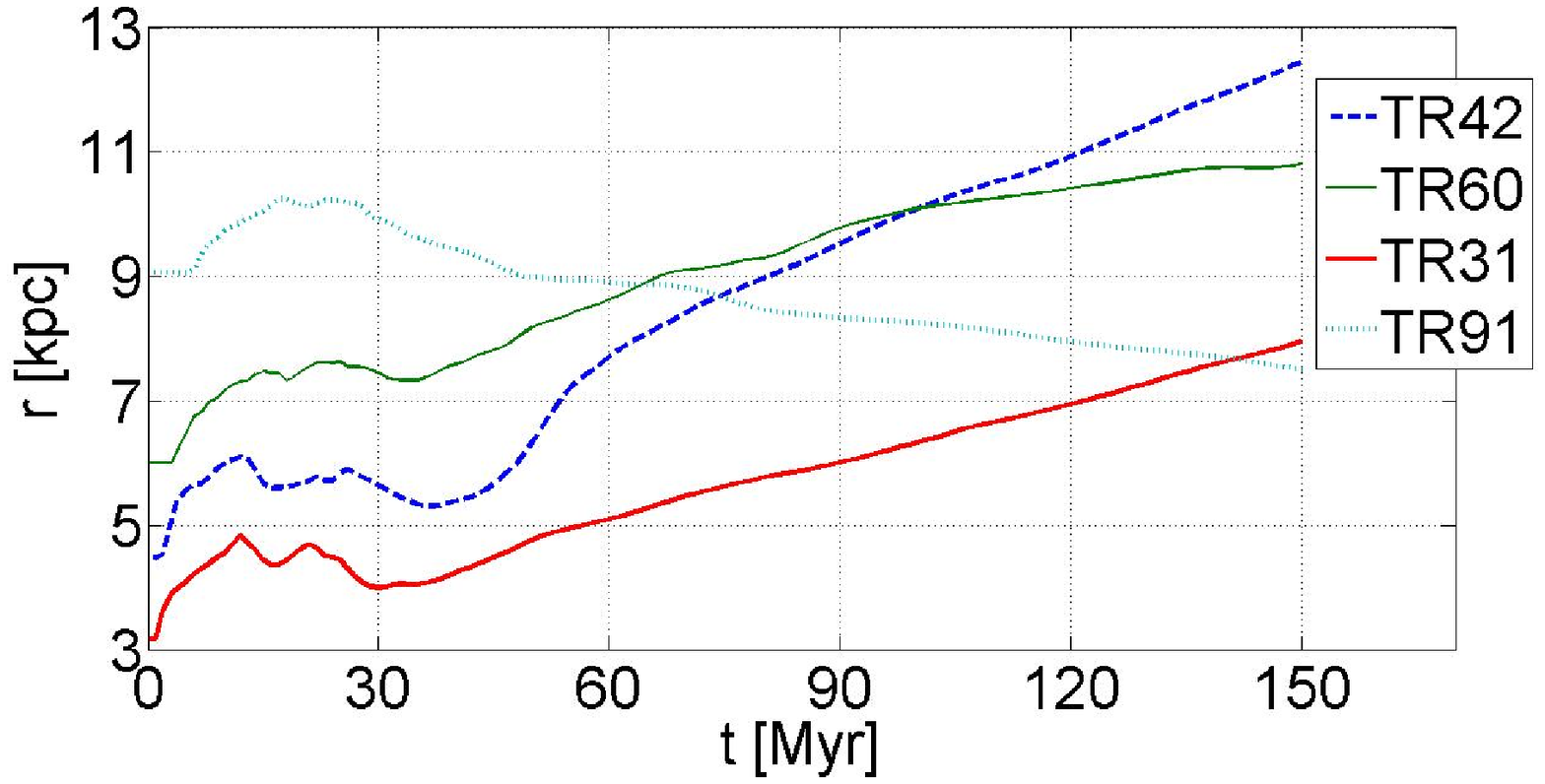}}
      \caption{Distance from center of the centers of mass of four traced regions as function of time.
      As before, TR31 stands for the tracer that started from the location $(\varpi, z)_0=(3,1)\kpc$, etc.
      The traced regions start to move outward when the forward shock hits them.
      At later times some of the tracers lose their shapes, and mixing over an extended volume determines the location of the center of mass
      of each tracer.              }
      \label{fig:rtracers}
\end{figure}

Another effect of the jets on the ICM is the displacement of the ICM.
We follow the center of mass of a tracer by taking the quantity $Q_i$ in equation (\ref{eq:tracers1}) to be the location
of the material $Q_i \rightarrow \vec{r_i}$.
In figure \ref{fig:rtracers} we present the distance $r$ from the center of the centers of mass of four tracers.
All tracers are pushed outward at early time as the forward shock passes through them.
Later they can fall back intact, as TR91 does, or be completely mixed, a process that can cause the center
of mass distance from the center to increase or decrease several times.

The mixing of material from different regions driven by vortices may affect the metallicity
gradient in the ICM. In order to properly examine this, the simulations need to include realistic star formation and metal
enrichment due to it, and more important, to include metallicity-dependent cooling.
The reason for the latter is that higher metallicity regions will in general cool faster and fall back to the center.
This is beyond the scope of the present work, and we leave it for a future work.

\section{HEATING MATERIAL NEAR THE EQUATORIAL PLANE}
\label{sec:heating}

\begin{figure}
  \centering
    {\includegraphics[width=0.7\textwidth]{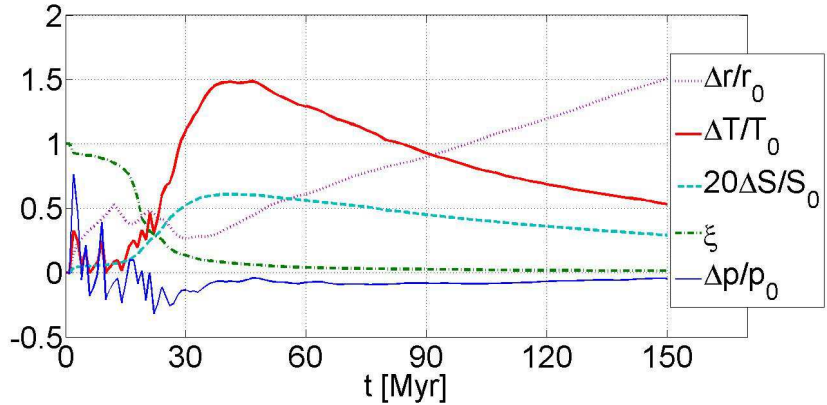}}
      \caption{Variation of the relative change in average pressure, temperature, location, entropy, and the mixing degree as defined in
      equation (\ref{eq:tracers3}), of the traced region TR31.
      Both temperature and entropy clearly show that the major heating process is mixing, indicated by decreasing value of $\xi$,
      and not the forward shock that hits the boundary of the tracer at $t= 1 \Myr$.
      The decrease in temperature and entropy at late times is due to mixing with cooler ICM regions.
      The cooling of the tracer comes with heating more ICM. }
      \label{fig:figtr31}
\end{figure}

To better understand the heating processes of the ICM gas residing near the equatorial plane we follow the variation
of some thermodynamic quantities of several traced regions.
In figure \ref{fig:figtr31} we present the relative changes of the average pressure, temperature, location, entropy, and the mixing degree of TR31;
TR31 is a traced region whose cross section in the $(\varpi, z)$ at $t=0$ is a circle of radius $0.25 \kpc$ centered on $(\varpi, z)_0=(3,1)$.
The average pressure, temperature, and location are calculated by equation (\ref{eq:tracers1}), and $\Delta Q \equiv Q_{ab}(t)-Q_{ab}(t=0)$.
The entropy (per particle) is calculated from the Sackur-Tetrode equation, using the average temperature and average density of the tracer.
Also shown is the degree of mixing, which is defined using equation (\ref{eq:tracers1}), so that
\begin{equation}
\xi \equiv \frac{\Sigma_i \xi_i^2 M_i}{\Sigma_i \xi_i M_i}.
\label{eq:tracers3}
\end{equation}

As can be seen in figure \ref{fig:figtr31}, initially $\xi=1$, which means there is no mixing.
When the shock passes through the region, there is a slight decrease in $\xi$. This is a numerical effect - as the region shifts its location,
the zones on the circumference appear to be mixed, although the region stays intact. At later times the value of $\xi$ further decreases due to mixing.

\begin{figure}
  \centering
    {\includegraphics[width=0.7\textwidth]{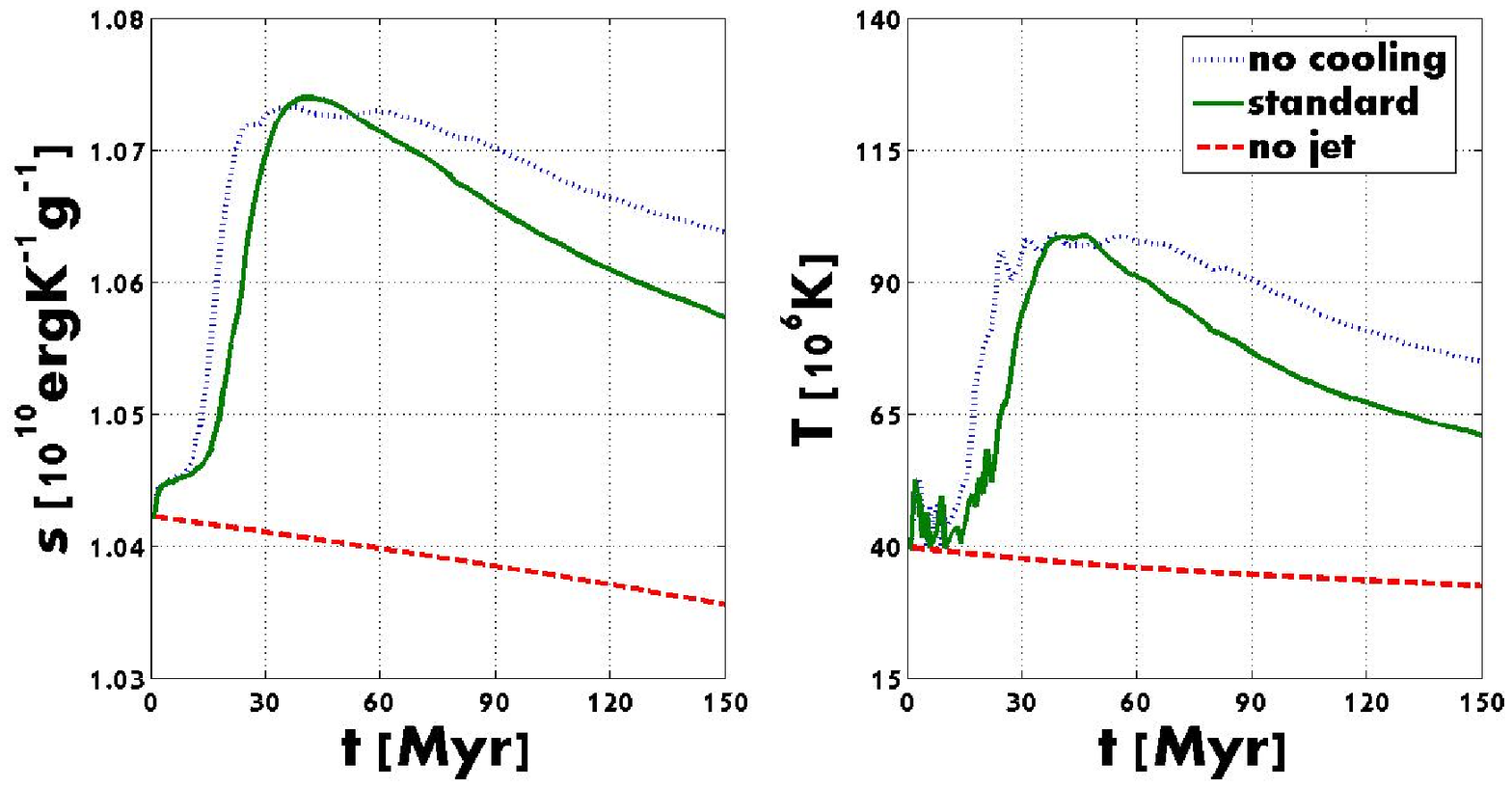}}
      \caption{Average entropy (left) and temperature (right) of TR31 for three cases as marked in the inset:
      the standard case where both the jet and radiative cooling are included, an active jet included but radiative cooling is not included, and
      a case with radiative cooling but no jet.
      The initial cooling time of the TR31 gas is $\tau_{\rm cool} \equiv (5/2)nkT/ n_e n_p \Lambda \simeq 6 \times 10^8 \yr$.
      The main differences between the standard run and that without radiative cooling is in the mixing process, where in the latter case
      the ICM is somewhat hotter at late times.
       }
      \label{fig:figcompare31}
\end{figure}

The main conclusion from figure \ref{fig:figtr31} is that the heating is due to mixing.
The forward shock that runs through the ICM has no significant lasting influence.
The forward shock hits TR31 at $t=1 \Myr$, compresses it and heats it.
However, the gas re-expands, and at $t \sim 10 \Myr$ its temperature returns to its original value.
Its entropy is somewhat higher than its original value, but not by much (see eq. 3.7 in \citealt{Soker2001}).
A significant heating of the TR31 gas starts at $t \sim 15 \Myr$ when the degree of mixing with hot gas increases
(as seen by the decreasing value of $\xi$).
The mixing is with hot shocked jets' and ICM gas; the ICM and the jet materials are already mixed as can be seen in figure \ref{fig:mixing1}.
At $t \ga 50 \Myr$ the temperature drops because mixing with cooler ICM medium starts to be more important than mixing with hot
shocked jet's gas.
In figure \ref{fig:figcompare31} the evolutions of the entropy (left) and temperature (right) of TR31 with
jet and radiative cooling included (the standard run) are compared to cases where either the jet or radiative cooling are not included.
Because of the stochastic nature of the mixing process, small differences in the initial conditions can lead to local differences
in the thermal evolution. The differences at early times between the standard simulation and the one with no radiative cooling
is due to this stochastic behavior. At late times the tracer TR31 in the standard simulation is cooler than the case with no radiative cooling.

\begin{figure}
  \centering
    {\includegraphics[width=0.7\textwidth]{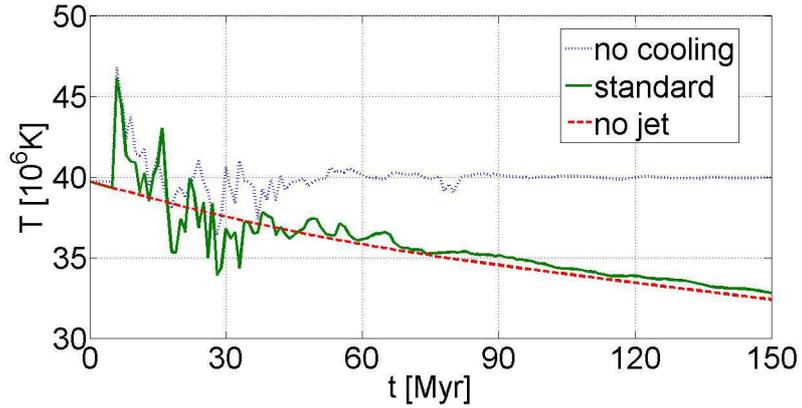}}
      \caption{Like figure \ref{fig:figcompare31} for the temperature, but for tracer TR91.
       }
      \label{fig:figtr91}
\end{figure}

Figure \ref{fig:figtr91} is similar to figure \ref{fig:figcompare31} but for tracer TR91
and only for the temperature.
This tracer stays intact till the end of the simulation ($t=150 \Myr$), although it loses its circular cross section.
It suffers a small degree of mixing, which accounts for it being somewhat hotter at late times than the case where there is no jet.
The rapid temperature rises of TR91 are due to the forward shock and sound waves that cross the TR91 material.
After the passage of each sound wave the TR91 gas cools back.
The average temperature of the TR91 gas is below its initial temperature, but somewhat above the case without jet activity.
This shows that the efficiency of heating decreases further away from the center.
The heating of this region will either take place if the jets of the next episode will have a different direction, such that
TR91 will be closer to the jets' axis, or if the next activity episode will occur after $\sim 100-300 \Myr$.
In this latter possibility the material of TR91 will be closer to the center (see figure \ref{fig:rtracers}),
and will be more likely to be mixed with the hot bubble material.
At $t=150 \Myr$ the center of mass of TR91 is at a distance of $7.5 \kpc$ from the center (fig. \ref{fig:rtracers}),
and falling slowly inward at a velocity of $\sim 0.01 \kpc \Myr^{-1}$.
At this rate, by $t \sim 300 \Myr$ TR91 will be close to where TR60 was initially,
close enough to the center to be heated by mixing (see below).
In any case, in the cold feedback mechanism some of the gas does cool, falls inward and feeds the AGN. So there is no need for 100\% heating efficiency. 

\begin{figure}
  \centering
    {\includegraphics[width=0.7\textwidth]{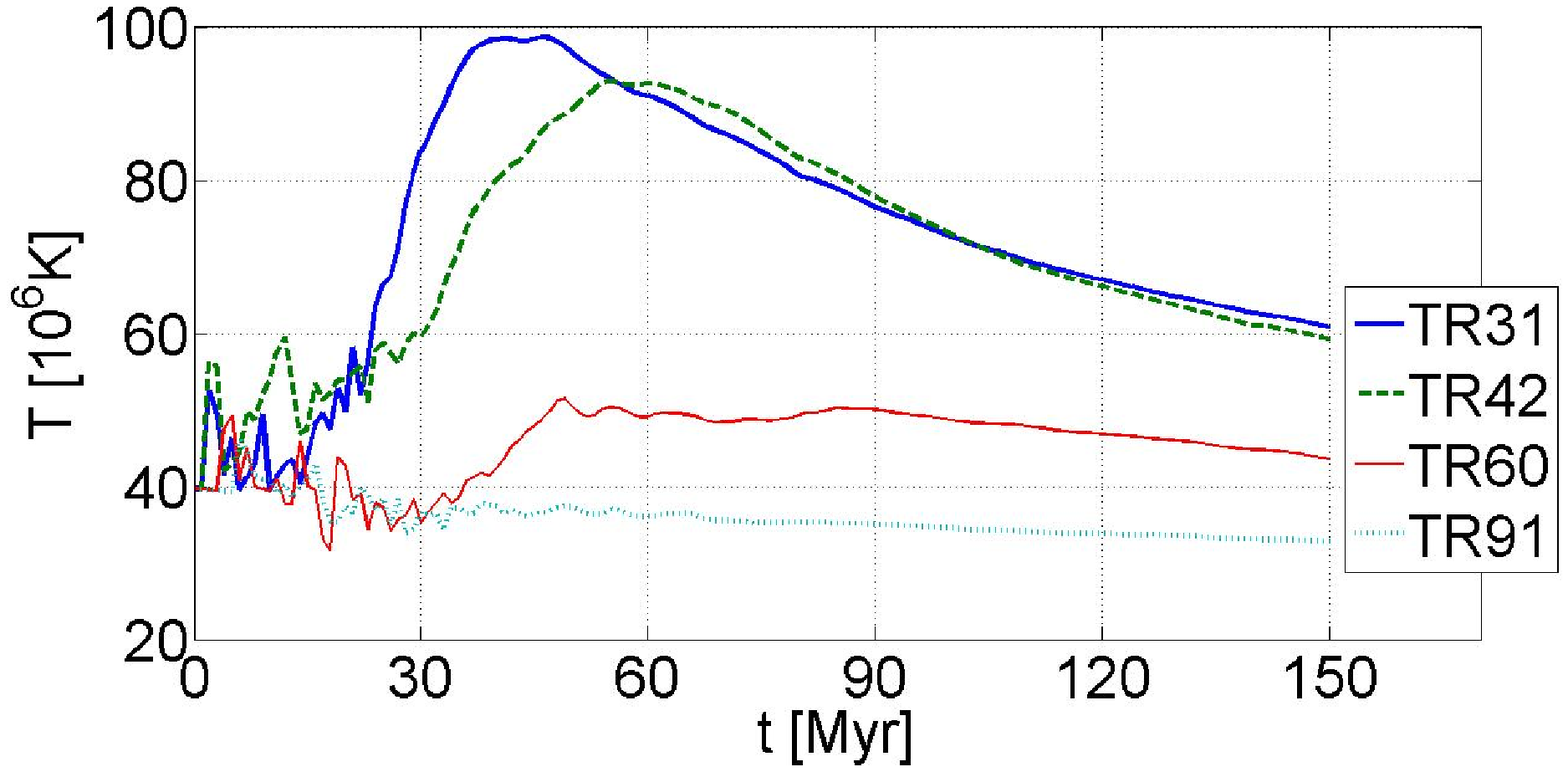}}
      \caption{Average temperature of four tracers as function of time. }
      \label{fig:Ttracers}
\end{figure}

\begin{figure}
  \centering
    {\includegraphics[width=0.7\textwidth]{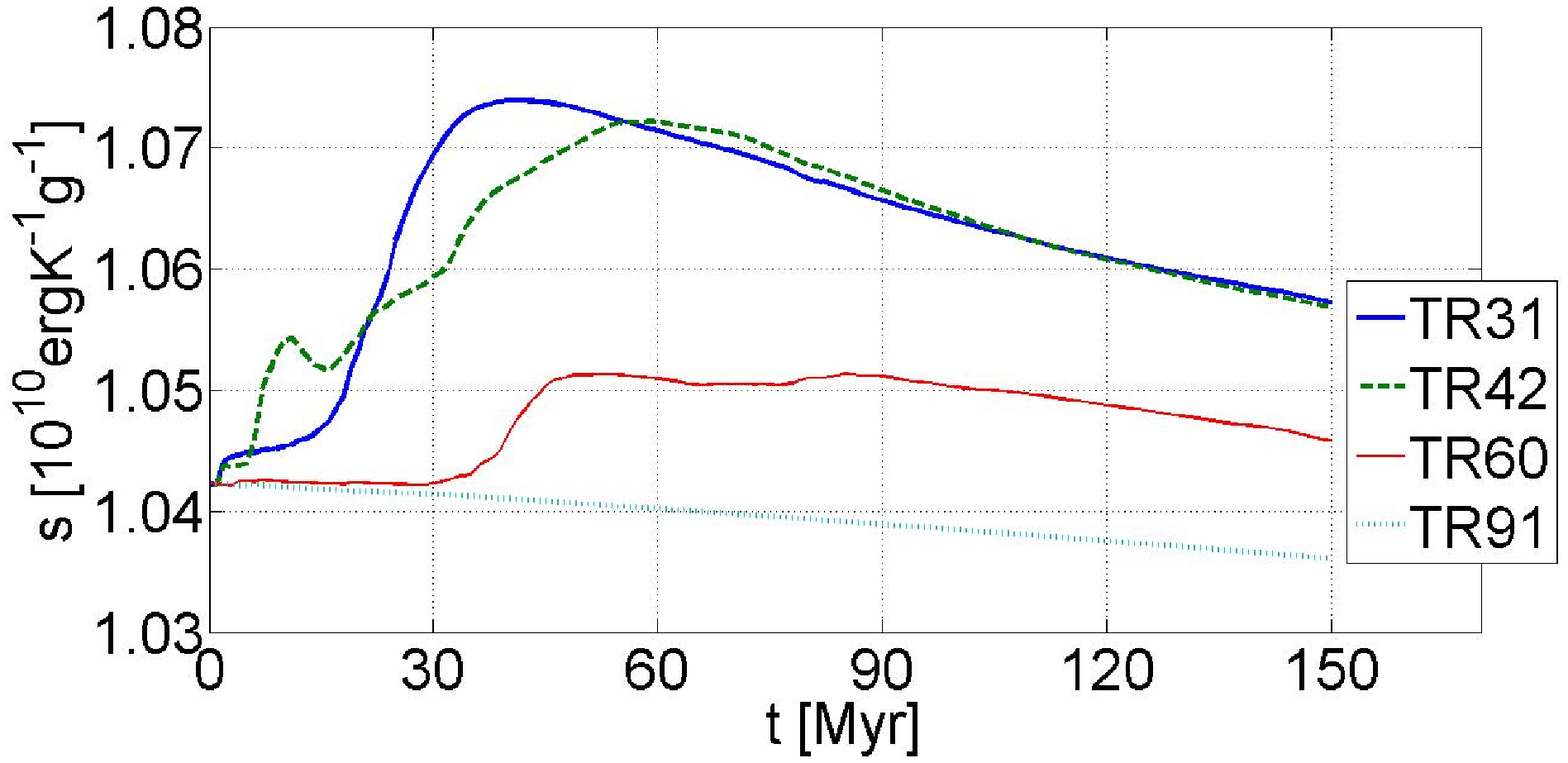}}
      \caption{The evolution of the specific entropy of four tracers as marked in the inset. The entropy of each tracer is calculated
      from its average density and temperature. }
      \label{fig:stracers}
\end{figure}

Figures \ref{fig:Ttracers} and \ref{fig:stracers} depict the time evolution of the average temperature and entropy of several traced regions.
All regions undergo shock heating early on, with diminishing shock heating as the region is further away from the center
and at a larger angle to the jets' direction.
However, shortly afterwards the regions cool down due to adiabatic expansion.
Subsequent heating is not by shocks, but by mixing of the traced region material with hot material from the shocked jet and ICM.
Mixing is due to the vortices and turbulent nature of the jet-ICM interaction process.

\section{SUMMARY}
\label{sec:summary}
We used the PLUTO hydrodynamic code \citep{Mignone2007} to conduct 2.5D hydrodynamic simulations, i.e., the flow is 3D but with an imposed
azimuthal symmetry around the $\theta =0$ ($z$) axis, to study the heating of gas perpendicular to the jets' axis.
Only one side of the equatorial plane was simulated, and the jet was active for a time period of $20 \Myr$.
We marked several intra-cluster medium (ICM) regions, the traced regions, and followed the evolution of their averaged thermodynamic variables.
Each traced regions has an initial circular cross section with radius of $0.25 \kpc$ in the meridional plane ($\varpi, z$), i.e., it is a torus in 3D.
At late times the traced region are vigorously mixed with the ICM and lose their intact structure.

We reproduced (fig. \ref{fig:presentation}) the formation of a fat bubble by a slow massive wide (SMW) jet \citep{Sternberg2007}, and the formation
of multiple sound waves with a single jet activity episode \citep{Sternberg2009}.
Vortices play major roles in the structure of evolution bubbles, as is evident also from figures \ref{fig:evolution} and \ref{fig:velocity}.
In the present paper we found that vortices play a major role in setting a complex flow structure that leads to a very efficient mixing of the
high entropy shocked jet gas with the ICM.
As evident from figure \ref{fig:mixing1} the mixing is very efficient both along the jet's expansion cone (white contours) and
near the equatorial plane (black contours).
The ICM gas along the jet's cone is dredge up to large distances, while the material near the equatorial plane
spreads and mixes in a large volume in the inner region.

Our main results are summarized in figures \ref{fig:figtr31} and \ref{fig:figtr91}.
These are that ($i$) the heating is very efficient near the equatorial plane up to a distance of $\sim 10 \kpc$,
($ii$) the heating by the forward shock wave is very small,
and ($iii$) that the main heating is caused by mixing; mixing is depicted by decreasing value of $\xi$.
These are further demonstrated in figure \ref{fig:figcompare31}, where the standard run is compared with similar simulations
but without either radiative cooling or without a jet.
The mixing and heating near the equatorial plane continues to counter radiative cooling for times of $\ga 10^8 \yr$ after the jets have ceased to exist.

Our results further have the following implications.
(1) The complex mixing will entangle magnetic field lines and will suppress any global heat conduction in the ICM near the center.
Namely, the presence of bubbles in cooling flows implies that heating the inner region by heat conduction is prohibited.
(2) The same entanglement process mixes the magnetic fields of the ICM and the shocked jets' material.
This will lead to reconnection of the magnetic field lines, hence will allow for local heat conduction between
the mixed ICM and jets' gas.
We emphasize the necessity to distinguish between the efficient process of local heat conduction (scales of $\la 0.1 \kpc$)
as opposed to the inefficient global (scales of $\ga 1 \kpc$) heat conduction process (see review by \citealt{Soker2010b}).
(3) The most crucial region for heating in cooling flows is the inner $\sim 10 \kpc$.
Our results imply that heating by jets that inflate bubble is very efficient in this inner region in all directions,
along and perpendicular to the jets' axis.
(4) The vigorous mixing implies that the region within few$\times 10 \kpc$ is multi-phase.
Some small regions will be the coolest ones. If they are not heated by another jet activity episode,
they will cool, flow inward, and feed the AGN.
Our results therefore supply further support to the cold feedback mechanism as suggested by \cite{Piz2005}.
The cold feedback mechanism has been strengthened recently by observations of cold gas and by more detailed studies (e.g.,
\citealt{Revaz2008, Pope2009, Wilman2009, Piz2010, Wilman2011, Nesvadba2011, Cavagnolo2011, Gaspari2012a, Gaspari2012b,
McCourt2012, Sharma2012, Farage2012}).

The same feedback mechanism that works in group and cluster cooling flows can be the feedback
mechanism during galaxy formation if the ISM was hot, i.e., at about the virial temperature there.
Namely, a cooling flow might have existed during galaxy formation periods \citep{Soker2010a}.
We note though that at the time of galaxy formation the  volume outer to the inner $\sim 10 \kpc$ contains much less
mass and its pressure is much lower than the values simulated here.
This implies the following. All traced regions here are accelerated by the forward shock to velocities of
$\sim 100-300 \km \s^{-1}$, and can be further pushed out by the shocked jets' gas.
However, they slow down because of the interaction with the ICM further out, as evident from figure \ref{fig:rtracers}.
During galaxy formation, on the other hand, this interaction can expel a large fraction of the gas outward.
The study of this type of interaction during galaxy formation is a subject of a future paper.

We thank an anonymous referee for helpful comments.
This research was supported by the Asher Fund for Space Research
at the Technion, and the Israel Science foundation.

\end{document}